\documentclass[prb,twocolumn,showpacs,epsfig,longeq]{revtex4}
\usepackage{graphicx}
\usepackage{dcolumn}
\usepackage{amssymb}
\usepackage{bm}
\usepackage{epsfig}
\usepackage{psfrag}

\def\3dots{\:\raisebox{-0.5ex}{$\stackrel{\textstyle.}{:}$}\:}
\def\beq{\begin{equation}}
\def\eeq{\end{equation}}
\def\bea{\begin{eqnarray}}
\def\eea{\end{eqnarray}}

\begin{document}

\title{Symmetry-dependent phonon renormalization in monolayer MoS$_2$ transistor}

\author{Biswanath Chakraborty$^1$}
\author{Achintya Bera$^1$}
\author{D. V. S. Muthu$^1$}
\author{Somnath Bhowmick$^2$}
\author{U. V. Waghmare$^2$}
\author{A. K. Sood$^1$}\thanks{Corresponding author}
\email{asood@physics.iisc.ernet.in}

\affiliation{$^1$Department of Physics, Indian Institute of Science, Bangalore - 560012, India}
\affiliation{$^2$ Theoretical Sciences Unit, Jawaharlal Nehru Centre for Advanced Scientific Research, Bangalore-560064, India}

\begin{abstract}

Strong electron-phonon interaction which limits electronic mobility of semiconductors can also have significant effects on phonon frequencies. The latter is the key to the use of Raman spectroscopy for nondestructive characterization of doping in graphene-based devices. Using in-situ Raman scattering from single layer MoS$_2$ electrochemically top-gated field effect transistor (FET), we show softening and broadening of A$_{1g}$ phonon with electron doping whereas the  other Raman active E$_{2g}^{1}$ mode remains essentially inert. Confirming these results with first-principles density functional theory based calculations, we use group theoretical arguments to explain why A$_{1g}$ mode specifically exhibits a strong sensitivity to electron doping. Our work opens up the  use of Raman spectroscopy in probing the level of doping in single layer MoS$_2$-based FETs, which have a high on-off ratio and are of enormous technological significance.

\end{abstract}


\maketitle

Discovery of graphene~\cite{NovoselovScience} stimulated an intense research activity due to interesting fundamental phenomena it exhibits as well as the techonological promise it holds in a broad range of applications ranging from sensors to nano-electronics. Vanishing bandgap of a single layer graphene is a sort of a limitation in developing a graphene-based field effect transistor with a high on/off ratio. This has spurred efforts to modify graphene to open up a gap and towards development of other two dimensional materials like MoS$_2$, WS$_2$ and boron nitride (BN), both experimentally and theoretically. Avenues to
open up gap through modification of graphene include quantum confinement in nanoribbons~\cite{AvourisRibbon}, surface functionalization ~\cite{ACSRoche}, applying electric field in the bilayer \cite{Morpugo,BisNano}, deposition of graphene on other substrates like BN ~\cite{PRBTheory,Ajayan}, and B or N substitutional doping~\cite{CNRAdv}, which require fine control
over the procedure of synthesis. 

In contast to graphene, single layer MoS$_2$ consisting of a hexagonal planar lattice of Mo atoms sandwiched between two similar lattices of S atoms (S-Mo-S structure) with intralayer covalent bonding is a semiconductor with a direct band gap of $\sim$ 1.8 eV, and is quite promising for FET devices with a high on-off ratio. It has been shown that the luminescence quantum yield of monolayer MoS$_2$ is higher than its bulk counterpart~\cite{Pl,Heinz}.Recently a monolayer  MoS$_2$ transistor~\cite{Transistor} has been shown to exhibit an on-off ratio of $\sim$10$^8$ and electron mobility of $\sim$200 cm$^2$/V-sec. These values are comparable to silicon based devices  and make MoS$_2$ based devices worth exploring further. It is known that in a field effect transistor, carrier mobility is limited by scattering from phonons and the maximum current is controlled by hot phonons. Both these issues in a FET depend on the electron-phonon coupling (EPC). Raman spectroscopy has been very effective to probe EPC for single~\cite{Ferrari,KimSLG,AndyNat} and bilayer graphene ~\cite{KimBLG,AndyBi,PimentaBi} transistors by investigating the renormalization of the G and 2D modes as a function of carrier density.

\begin{figure}[htbp]
\includegraphics[width=0.45\textwidth]{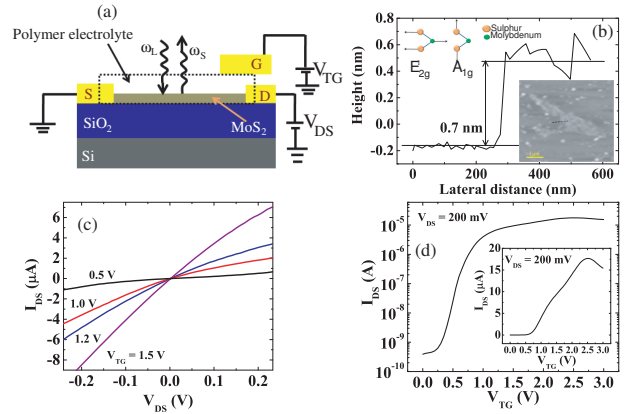}
\caption{\label{1}(Color online) (a) Schematic of experimental set up. Source, drain and gate electrodes were marked as S, D and G respectively. The dotted box is the polymer electrolyte layer. (b) AFM height profile of the monolayer MoS$_2$ flake. (Right Inset) AFM image showing the dotted line along which the height profile is taken.  Scale bar is 1 $\mu$m. (Left Inset) Atomic displacements, indicated by arrows, corresponding to E$_{2g}^{1}$ and A$_{1g}$ modes as viewed along the [1000] direction. (c) I$_{DS}$-V$_{DS}$ plot at various top gate voltage V$_{TG}$. (d) I$_{DS}$ as a function of V$_{TG}$ at V$_{DS}$ = 200 mV. The on-off ratio was $\sim$ 10$^5$ and the device mobility was estimated to be 50cm$^2$/V-sec. Inset shows the transfer characteristic in linear scale.  }
\end{figure}

Recent layer-dependent Raman studies of single and few layers of MoS$_2$~\cite{Raman} have shown that freuqency of E$_{2g}^{1}$ phonon increases as the number of layers decreases, whereas the frequency of A$_{1g}$ phonon decreases~\cite{Raman}. This has been recently explained in terms of enhanced dielectric screening of the long range Columb interaction between the effective charges with         increasing  number of layers~\cite{SanchezPRB}. For non-destructive characterization of carriers in recently developed mono-layer MoS$_2$ transistor~\cite{Transistor}, Raman spectroscopy can be quite useful and requires precise knowledge and understanding of phonon renormalization of single MoS$_2$ layer as a function of carrier concentration. The temperature dependence of the mobility in n-type bulk MoS$_2$ had been attributed to the scattering of carriers by optical phonons that modulate thickness~\cite{HallMobility} implying that the A$_{1g}$ phonons with atomic displacements parallel to the c-axis should be involved in controlling mobility of the carriers.

In this letter, we report in-situ carrier dependent Raman study of a top gated single layer MoS$_2$ transistor achieving a maximum electron doping of $\sim$2$\times$10$^{13}$/cm$^2$. The transfer characteristic of the top gated device shows an on-off ratio $\sim$ 10$^5$ and a field effect mobility of 50 cm$^2$/V-sec. We show that the A$_{1g}$ mode shows a strong doping dependence; the phonon frequency decreases by 4 cm$^{-1}$ and linewidth broadens by 6 cm$^{-1}$ for electron doping of 1.8 $\times$10$^{13}$/cm$^2$. Phonon frequency and linewidth of the E$_{2g}^{1}$ mode show a much less dependence on carrier concentration. The difference in the behaviour of these optical phonons are explained quantitatively using density functional theory (DFT).

Fig.~\ref{1}a shows a schematic of our experimental set up. Single layer MoS$_2$ flakes were mechanically exfoliated from a bulk single crystal procured from M/s. SPI Supplies and transferred on a 300 nm SiO$_2$ on a degenerately doped p-type silicon substrate (procured from M/s. XT Wafer). After optical identification, the flake height is measured by contact mode AFM to be 0.7 nm, in  agreement with the S-Mo-S layer thickness (see Fig.~\ref{1}b). Standard electron-beam lithography and deposition were done to form $\sim$50 nm thick Au contacts as source (S) , drain (D) and gate (G) electrodes. Room temperature Raman spectra were recorded with 514.5 nm laser excitation with Witec confocal spectrometer using 50X long working distance objective. Laser power was kept below 1mW to avoid sample heating. Electrical measurements were done with Keithley 2400 source meters. For top gating, we have used solid polymer electrolyte comprising of a mixture of LiClO$_4$ and polyethylene oxide (PEO) in the weight ratio 1:8. Solid polymer electrolyte as the gate material was chosen because of its high capacitance ~\cite{BisNano} enabling high carrier concentration with low gate voltage ($\leq$~ 2V). At the same time, being almost transparent, it allows us to perform in-situ optical measurements with simultaneous electrical characterization. Fig.~\ref{1}c shows drain-source current (I$_{DS}$) as a function of drain-source bias (V$_{DS}$) for a representative device of length (L) 2.5 $\mu$m and width (W) 1.5 $\mu$m. A small non linearity may be  due to the Schottky barrier at the contacts. The transfer characteristics of the device in Fig.~\ref{1}d is plotted in a semi log scale. The on-off ratio for our device is $\sim$10$^5$ and  the low field effect mobility ($\mu$) of our device was calculated ($\mu=\frac{L}{W}\frac{g_m}{V_{DS} C_{TG}}$) to be 50 cm$^2$/V-sec. Here g$_m$ is the transconductance (g$_m=\frac{\partial I_{DS}}{\partial V_{TG}}$) of the device. Enhanced photoconductivity was observed when the incident laser beam was focused on the sample at gate voltages of 0.5, 1.0, 1.5 and 2.0V in agreement with recent reports~\cite{Heinz}. The photoconductive response was large at lower gate voltages but decreases at higher gate voltages when the conduction band starts getting populated. Raman spectrum at each gate voltage was recorded only after the stabilization  of the channel current. The threshold gate voltage V$_T$, at which the device switches from `off' to `on' state was $\sim$0.1 V. The gate induced electron concentration \textit{n} is estimated using $ne$ = C$_{TG}(V_{TG} - V_T$). The value of C$_{TG}$ is 1.5 $\mu$F/cm$^2$. In brief, we extracted the value of top gate capacitance C$_{TG}$ from an experiment with bilayer graphene transistor involving dual gate configuration ~\cite{BisNano}. The back gate dielectric was 300 nm SiO$_2$.  Several back gate sweeps were done for fixed top gate voltages. Since a maximum in the resistance corresponds to the charge neutrality point,  the carrier concentration induced by both the gate voltages were equated to get the C$_{TG}$ to be 1.5 $\mu$F/cm$^2$. Since the electrolyte in the present experiments is same as in Ref.[5], we take this value in estimating $n$.

\begin{figure*}[htbp]
\includegraphics[width=0.9\textwidth]{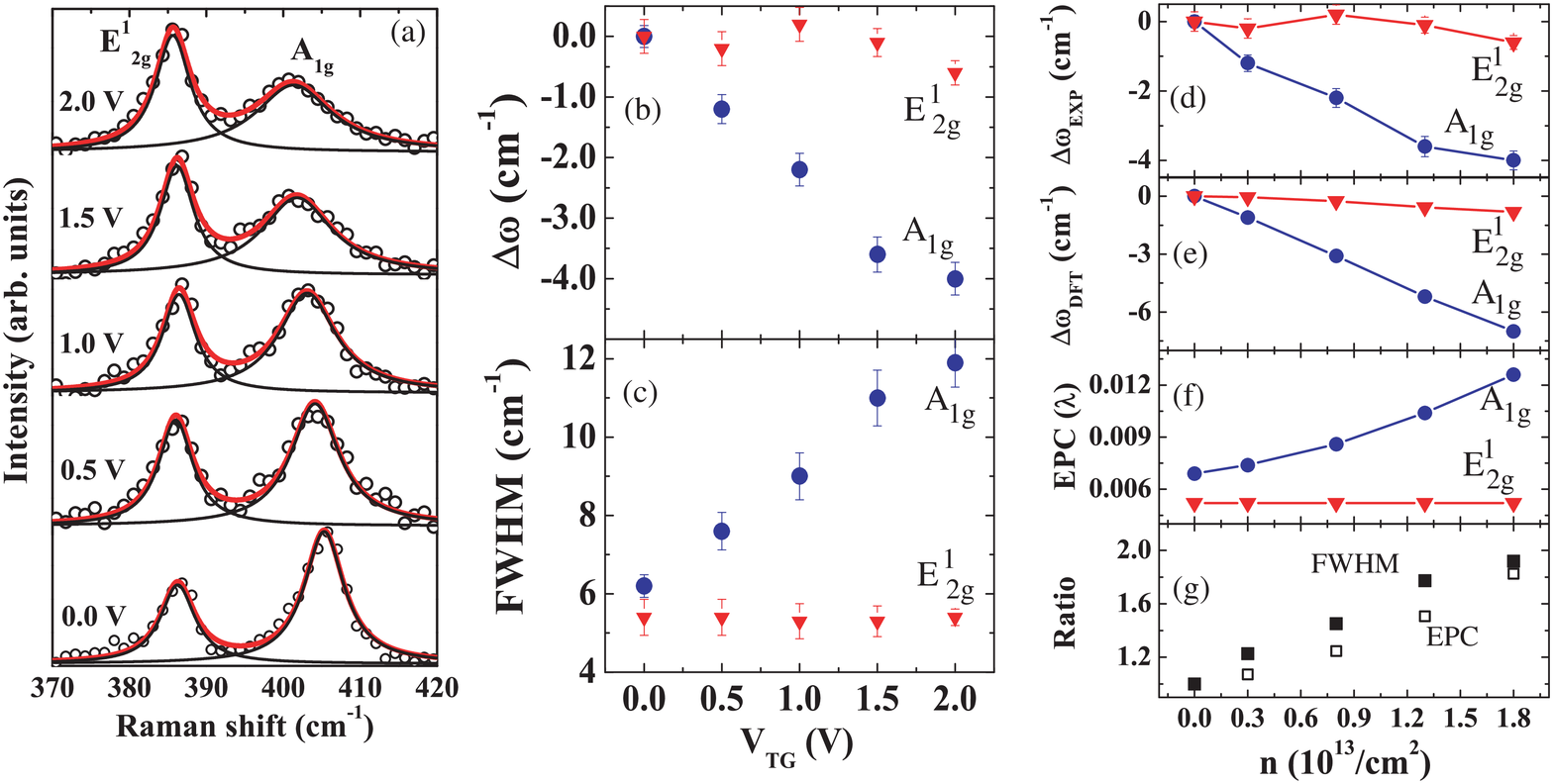}
\caption{\label{2}(Color online) (a) Raman spectra of monolayer MoS$_2$ at different top gate voltages V$_{TG}$. Open circles are experimental data points, the gray (red) lines are Lorentzian fits to the total spectrum and the black color lines are the Lorentzian fit to individual peak. Change in the (b) phonon frequency $\Delta\omega$ and (c) FWHM of A$_{1g}$ and E$_{2g}^{1}$ modes as a function of V$_{TG}$. Change in zone center phonons $\Delta\omega$ of (d) from experiment  and (e) from DFT calculations as a function of electron concentration $n$. (f) Electron-phonon coupling of A$_{1g}$ and E$_{2g}^{1}$ modes as a function of $n$. (g) Ratio of EPC [$\lambda_{A_{1g}}(n\neq0)$/$\lambda_{A_{1g}}(n=0)$] shown by open squares and phonon linewidth [FWHM$_{A_{1g}}(n\neq0)$/FWHM$_{A_{1g}}(n=0)$] shown by filled squares as a function of $n$. }
\end{figure*}

Fig.~\ref{2}a shows the evolution of zone center phone  E$_{2g}^{1}$ and A$_{1g}$ modes  of MoS$_2$ monolayer at different top gate voltages. As depicted in Fig.~\ref{1}b inset, A$_{1g}$ phonon involves the sulphur atomic vibration in opposite direction along the c axis (perpendicular to the basal plane) whereas for E$_{2g}^{1}$ mode, displacement of Mo and sulphur atoms are in the basal plane. Lineshape parameters were obtained by fitting a sum of two Lorentzians to the data. Fig.~\ref{2}b and Fig.~\ref{2}c show the shift of the mode frequencies and the corresponding full width at half maximum (FWHM), respectively, as a function of gate voltage. The dependence of the change in mode frequencies [$\Delta\omega = \omega(n \neq0) - \omega(n=0)$] on the carrier concentration ($n$)  is shown in Fig.~\ref{2}d. For a maximum electron concentration of 1.8 $\times$ 10$^{13}/cm^2$, the A$_{1g}$ mode frequency softens by 4 cm$^{-1}$, as compared to only $\sim $ 0.6 cm$^{-1}$ for the E$_{2g}^{1}$ mode. The linewidth of the A$_{1g}$ mode increases significantly by $\sim$6 cm$^{-1}$ for the maximum doping achieved, whereas the linewidth of the E$_{2g}^{1}$ mode does not show any appreciable change. These results show that the A$_{1g}$ phonon renormalization could be used as in-situ read-out of the carrier concentration in MoS$_2$ devices. We will now quantitatively understand the  different renormalization of the two modes, A$_{1g}$ and E$_{2g}^{1}$ due to electron-phonon interaction.

Our calculations are based on first-principles density functional theory as implemented Quantum Espresso package,~\cite{T1} a plane-wave basis set (70 Ry cutoff) and norm conserving pseudopotentials (Rappe-Rabe-Kaxiras-Joannopoulos~\cite{T2}). Exchange-correlation energy of electrons is approximated with a local density approximation and a parametrized functional of Perdew and Zunger~\cite{T3}. A monolayered form of $\textrm{MoS}_2$ is simulated using a periodic supercell, with a vaccum of $\sim 15$ \AA~ separating adjacent periodic images along the $z$ direction. Integrations over the Brillouin zone were sampled with uniform $24\times 24\times 1$ and $48\times 48\times 1$ $k-$point meshes in calculation of total energy and electron-phonon coupling, respectively.

\begin{figure}[htbp]

\includegraphics[width=0.45\textwidth]{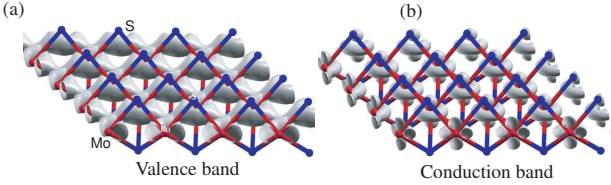}

\caption{\label{5}(Color online) Contour plots (in light gray) of charge density  ($|\psi(r)|^2$)
(a) at the top of the valence, and (b) bottom of the conduction bands at $K$ point.
While states near the top of valence band consist of Mo $4d$  states with some 
hybridization with S $3p$ states, those near the conduction band edge have clearly the character of d$_{z^2}$ state of Mo. The dark gray (red) and black (blue) spheres are Mo and sulphur atoms respectively.   }
\end{figure}

Our calculations show that monolayer $\textrm{MoS}_2$ is a direct band gap semiconductor with a band gap of 1.8 eV at K-point, in good agreement with known results~\cite{T4}. Valence and conduction band edges primarily consists of  Mo $4d$ states with some hybridization with S $3p$  states [see Fig.~\ref{5} a and b]. In particular, states near the bottom of conduction band near the K-point have a character of d$_{z^2}$ state of Mo, and that at the top of valenec band has d$_{xy}$ character. We note that charge density ($|\psi(r)|^2$) associated with each of these states has a full symmetry of MoS$_2$ layer, which has important consequences for electron-phonon coupling (to be elaborated later).

We simulated electron doping in MoS$_2$ by adding a small fraction of electrons to its unit cell. Doping has contrasting effects on the frequencies of A$_{1g}$ and E$_{2g}^{1}$ optic modes as shown in Fig.~\ref{2}e. While the former mode is found to soften significantly ($\sim 7 ~\textrm{cm}^{-1}$ at $\sim 1.8\times 10^{13}/\textrm{cm}^2$ doping), the latter is hardly affected, in very good agreement with our experimental results shown in Fig.~\ref{2}d. In order to understand this trend, we carry out a systematic study of electron-phonon coupling as a function of electron doping.  The electron-phonon coupling (EPC) of a mode $\nu$ at momentum $\mbox{\boldmath $q$}$ (with frequency $\omega_{\textbf{\textit{q}}\nu}$) is calculated as ~\cite{MauriNanoletters}

\begin{eqnarray}\nonumber 
\label{lambda}
\lambda_{\mbox{\scriptsize\boldmath $q$}\nu} &=&
\frac{2}{\hbar\omega_{\mbox{\scriptsize\boldmath $q$}\nu}N(\epsilon_f)}\sum_{\mbox{\scriptsize\boldmath $k$}}\sum_{mn}|g_{\mbox{\scriptsize\boldmath $k+ q,k$}}^{\mbox{\scriptsize\boldmath $q$}\nu ,ij}|^2\times\delta(\epsilon_{\mbox{\scriptsize\boldmath $k+ q$},i}-\epsilon_f) \\
&{\times}& 
\delta(\epsilon_{\mbox{\scriptsize\boldmath $k$},j}-\epsilon_f),
\end{eqnarray}
where $\omega$ and $N(\epsilon_f)$ is the phonon frequency and 
electronic density of states at the Fermi energy, respectively.
The electron-phonon coupling matrix element is given by

\begin{equation}
\label{g}
g_{\mbox{\scriptsize\boldmath $k + q,k$}}^{\mbox{\scriptsize\boldmath $q$}\nu ,ij}=\left(\frac {\hbar}{2M\omega_{\mbox{\scriptsize\boldmath $q$}\nu}}\right)^\frac{1}{2}\langle\psi_{\mbox{\scriptsize\boldmath $k + q$},i}|\triangle V_{\mbox{\scriptsize\boldmath $q$}\nu}|\psi_{\mbox{\scriptsize\boldmath $k$},j}\rangle,
\end{equation}
where $\psi_{\mbox{\scriptsize\boldmath $k$},j}$ is the electronic wavefunction with wavevector  $\mbox{\boldmath $k$}$  and energy eigenvalue $\epsilon_{\mbox{\scriptsize\boldmath $k$},j}$ for band $j$ and M is the ionic mass. $\triangle V_{\mbox{\scriptsize\boldmath $q$}\nu}$ is the change in the self-consistent potential associated with a phonon of wavevector $\mbox{\boldmath $q$}$, branch $\nu$ and frequency $\omega_{\mbox{\scriptsize\boldmath $q$}\nu}$. Eq~\ref{g} defines the scattering of an electron from band $j$ to band $i$ due to the phonon $\nu$ with momentum $\mbox{\boldmath $q$}$. Our results in Fig.~\ref{2}f show that A$_{1g}$ mode couples much more strongly with electrons than the E$_{2g}^{1}$ mode. This can be understood using group theoretical analysis of symmetry. A$_{1g}$ mode has a symmetry of the lattice (the identity representation, i.e. the structural distortions in this mode do not break the symmetry of MoS$_2$, see Fig.~\ref{1}b inset). As a result, all electronic states can have a nonzero expectation value in  Eq~\ref{g} for the perturbation of A$_{1g}$ mode, giving a large electron-phonon coupling in Eq~\ref{lambda}. Electron doping leads to occupation of the bottom of the conduction band at K-point states which have a character of d$_{z^2}$ of Mo [see Fig.~\ref{5}b]. The $|\psi(r)|^2$ of the states near $K-$point also transform according to the identity representation A$_{1g}$. Hence, changes in occupation of these states with electron doping yield a significant change in the EPC of the A$_{1g}$ phonon. In contrast, the matrix element in Eq~\ref{g} vanishes for E$_{2g}^1$ mode (orthogonality of A$_{1g}$ and E$_{2g}^1$ representations) and its coupling with electrons is weakly dependent on doping. Frey et al.~\cite{Frey} have put forward similar conclusions.

It is interesting to compare EPC of MoS$_2$ with that of graphene. First of all, electron doping of about 1.8x10$^{13}$ cm$^{-2}$ results in hardening of G-band of graphene ~\cite{AndyNat} by about 10 cm$^{-1}$ and softening of A$_{1g}$ mode of MoS$_2$ by about 4 cm$^{-1}$. The G-phonon renormalization occurs due to phonon induced electron-hole (e-h) pair creations. For the G mode (\textbf{q} $\sim$ 0) in graphene, it involves e-h creation within a valley. Doping the graphene blocks the generation of phonon induced e-h pairs and hence affecting the phonon self energy. Secondly, distortions of the structure with atomic displacements of a G-phonon lead to a mere shift of the centre of the Dirac cone (and in the Fermi-surface). If electrons follow these distortions remaining in their ground state (adiabatic limit), their energy cost does not change with doping and hence doping would not result in any shift of frequency of the G-band. However, it is the breakdown of adiabatic approximation (the fact that electrons do not follow nuclear motion remaining in their ground state) that is responsible for the energy cost and hardening of the G-band with electron and hole doping ~\cite{Ferrari}. Another way to explain is in terms of Kohn anomaly in graphene for \textbf{q} = 0 phonon which is weakened on doping. In contrast, in semiconducting MoS$_2$, phonon renormalization occurs within the adiabatic approximation. The electron doping results in occupation of the anti-bonding states in the conduction band of MoS$_2$ making the bonds weaker and the A$_{1g}$ mode, which preserves the symmetry of the lattice, softens. While the EPC of Raman mode at $\Gamma$ point of MoS$_2$ exhibits a strong dependence on doping, similar dependence is seen for $K-$point phonons of graphene~\cite{MauriNanoletters} and our symmetry-based explanation applied here as well. We note that the latter transforms according to identity representation of the symmetry group at (or near) the $K-$point. The EPC argument applies to the linewidth as well. Two different mechanism contribute to the phonon linewidth: (a) the EPC contribution (FWHM$^{EPC}$) and (b) the contribution (FWHM$^{an}$) arising from anharmonic effects~\cite{MauriEPCPRB,MauriEPCPRL}. We can express, FWHM = FWHM$^{EPC}$ + FWHM$^{an}$.  The phonon linewidth, FWHM$^{EPC}$, is proportional to EPC associated with a particular mode ~\cite{MauriEPCPRB,MauriEPCPRL}. The increase in A$_{1g}$ linewidth is a result of the strengthening of electron-phonon coupling ($\lambda$) with doping (Fig.~\ref{2}c and Fig.~\ref{2}f). Instead of the absolute values, we compare the ratios of FWHM (FWHM$_{n\neq0}$/FWHM$_{n=0}$) and EPC values ($\lambda_{n\neq0}$/$\lambda_{n=0}$) for the A$_{1g}$ mode (see Fig.~\ref{2}g). The FWHM ratio follows the same trend as the EPC ratio establishing that the increase in linewidth is due to increase in the electron-phonon coupling values ($\lambda$) with doping.

In summary, we have demonstrated that electron doping in single layer MoS$_2$ results in softening  specifically of its Raman active A$_{1g}$ phonon, accompanied by an increase in the line-width of its Raman peak. In comparison, the other Raman mode with E$_{2g}^{1}$ symmetry is quite insensitive to electron doping. This is due to a stronger electron-phonon  coupling of the A$_{1g}$ mode than of the E$_{2g}^{1}$ mode, confirmed with first-principles DFT calculations and symmetry arguments. Our work shows how Raman scattering can be effectively used to characterize level of doping in gated FET device based on single-layer MoS$_2$ with a high on-off ratio of $\sim$ 10$^5$ and having potential to be used in digital electronics and sensors. 
 

A.K.S. acknowledges the funding from Department of Science and Technology, India, under the Nanomission grant. U.V.W. acknowledges funding from AOARD grants FA2386-10-1-4062 and FA2386-10-1-4150 from US-Air Force. A.B. acknowledges the support as a CSIR fellow.

\end{document}